\documentclass[aps,amsmath,amssymb,twocolumn]{revtex4}
\usepackage{graphicx}
\usepackage{amsmath}

\begin{document}

\title{Method for Full Bloch-Sphere Control of a Localized Spin via a Single Electrical Gate}
\author{Joseph Pingenot}
\author{Craig E. Pryor}
\author{Michael E. Flatt\'e}
\affiliation{Optical Science and Technology Center and Department of Physics and Astronomy, University of Iowa, Iowa City, IA 52242
}

\begin{abstract}
We calculate the dependence on an applied electric field of the {\bf \emph{g}}~tensor of a single electron in a self-assembled InAs/GaAs quantum dot. We identify dot sizes and shapes for which one in-plane component of the {\bf \emph{g}}~tensor changes sign for realistic electric fields, and show this should permit full Bloch-sphere control of the electron spin in the quantum dot using only a static magnetic field and a single vertical electric gate.
\end{abstract}

\maketitle

Manipulating individual spins in solids requires quickly and coherently reorienting localized
spins while leaving neighboring spins
unaffected\cite{Awschalom2002}. 
Difficulties confining
oscillating magnetic fields have motivated alternate
approaches that use electric fields to change the local magnetic environment, including moving an electron within a hyperfine field gradient\cite{Petta2005} or fringe-field
gradient\cite{Tokura2006}.
Higher temperatures require spins to be localized in much smaller quantum dots, where these techniques are less effective. In contrast, approaches that couple the spin to an electric field via the spin-orbit interaction\cite{Kalevich1993,Oestreich1996b,Ivchenko1997b,Jiang2001,Salis2001,Kato2003,Doty2006,Nakaoka2007}, especially via {\bf \emph{g}}~tensor manipulation techniques\cite{Kato2003}, should be scalable to small dots with large confinement.
 Here we calculate the {\bf \emph{g}}~tensor of a single electron in a small quantum dot and show the symmetry of its electric field dependence permits full Bloch sphere control of the spin using an electric field applied in a single direction.
We find the spin manipulation frequency of an InAs/GaAs QD in 1 Tesla exceeds 150 MHz. 

The energy difference between a spin-up and spin-down state (Zeeman energy) of an electron in a zincblende direct-gap
semiconductor depends on the band gap and spin-orbit splitting in the
material\cite{Roth1959}. The Zeeman energy thus can be modulated in a semiconductor heterostructure, even for a static magnetic field, by changing the overlap of the electronic wave
function with different materials\cite{Ivchenko1997b,Jiang2001,Salis2001,Doty2006}. Other contributions in a system of low symmetry, such as a quantum well or quantum dot (QD), produce very different
Zeeman energies from the same magnetic field when it is applied
along different principal axes\cite{Ivchenko1997}, as
described by the Hamiltonian 
\begin{equation}
\label{eqn_zeeman}
H = -\frac{e}{2m_e}{\bf S \cdot \mathbf{g}\cdot B} = {\bf S \cdot \Omega}.
\end{equation}
Here $e$ is the electron charge, $m_e$ its mass,  {\bf
  S} is the electron spin operator, {\bf $g$} is the Land\'e {\bf g}
tensor, {\bf B} is the magnetic field, $\hbar$ is Planck's constant and ${\bf \Omega}$ is the spin
precession vector. From Eq.~(\ref{eqn_zeeman}), a change in the Zeeman energy in a static magnetic field corresponds to a change in the {\bf g} tensor.  As the {\bf g} tensor changes due to a changing applied electric
field in a semiconductor heterostructure, the spin precession vector ${\bf \Omega}$
changes both its magnitude and direction, permitting reorientation of the spin.  An oscillating electric field produces an oscillating  transverse ${\bf \Omega}$, which allows resonant control of the spin orientation in a static magnetic field, as demonstrated in quantum wells\cite{Kato2003}. As these manipulations only require electric-field modulation, traditional high-density electric gate design strategies may be applicable to high-density arrays of single-spin devices.

Several significant challenges stand in the way of manipulating single
spins using these methods. The strong confinement in QDs leads
to a quenching of the orbital motion of the electron, driving the
components of the {\bf g}~tensor towards the free-electron value of
$2$, and reducing the potential modulation of the {\bf g}~tensor in an
electric field\cite{Pryor2006}. The long spin coherence times for electrons relative to holes\cite{Meier1984,Kroutvar2004,Petta2005} favors their use as coherent single-spin entities. However, due to the greater effective confinement of electrons in a quantum dot, the electric-field tunability of the electron {\bf g} tensor would be much smaller than the corresponding tunability for holes or excitons.  Furthermore, to be able to
conveniently and rapidly reorient the spin to  any direction on
the Bloch sphere, the ability to generate precession around two
orthogonal axes is highly desirable, which might be expected to involve independent control of two electric fields. Manipulating an electric field along two directions in a nanoscale device would make the high-density integration of single-spin devices considerably more complex. We find that tall dots provide sufficient tuning for 150~MHz spin manipulation, and that full Bloch sphere control is possible with a single vertical gate.

\begin{figure}
\includegraphics[width=\columnwidth]{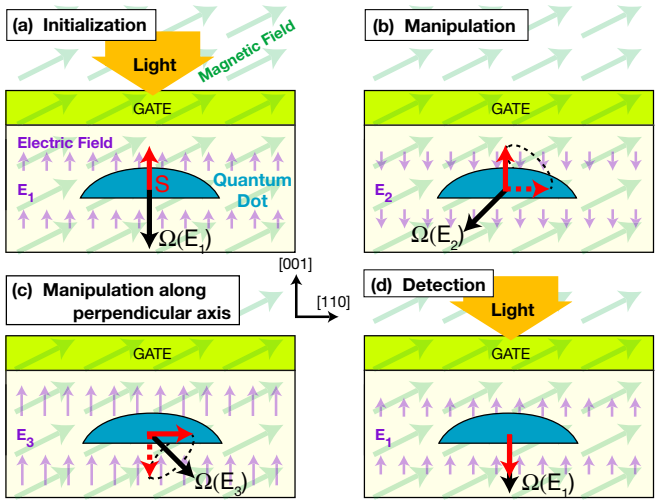}
\caption{
Sequence of applied electric fields for injection, manipulation, and detection of an electron spin in a single quantum dot.
(a)  for electric field $E_1$  the spin precession axis ${\bf \Omega}(E_1)$ is parallel to $[001]$, easing optical spin injection. (b)   for electric field $E_2$  the spin precession axis ${\bf \Omega}(E_2)$ rotates 45$^{\rm o}$ towards the [110] direction. (c) for electric field $E_3$  the spin precession axis ${\bf \Omega}(E_3)$ is perpendicular to ${\bf \Omega}(E_2)$. (d) return to an electric field $E_1$ eases optical measurement of the spin polarization along the [001] direction.\label{fig_cartoon}}
\end{figure}

We now consider a single electron in an InAs/GaAs dot in a static magnetic field, with a series of electric fields applied by a top gate which will re-orient the spin in an arbitrary direction (Fig. \ref{fig_cartoon}). 
The growth direction is [001] and the principal axes of the {\bf g}~tensor are [001], [110], and [1$\overline{1}$0].  
We will refer to the values of {\bf g} along these principal axes as the {\bf g} tensor components.  The magnetic field lies in the [001]--[110] plane, for reasons described later.
The initial field $E_1$ gives an ${\bf \Omega} ( E_1 )$ parallel to the [001] direction (Fig.~\ref{fig_cartoon}(a)), allowing efficient spin initialization by optical illumination\cite{Meier1984,Pryor2003}, and stabilizing the spin polarization until it is time to manipulate the spin. 
Guaranteeing this orientation of ${\bf \Omega}(E_1)$ for the magnetic field shown in Fig.~\ref{fig_cartoon} requires ${\rm g}_{\rm [110]}( E_1 )=0$ and ${\rm g}_{\rm [001]}( E_1 )\ne0$ (here assumed $<0$).
Figs.~\ref{fig_cartoon}(bc) show the  quantum dot's spin dynamics in response to the electric fields $E_2$ and $E_3$,  which produce spin precession vectors along two perpendicular axes.  
Fig.~\ref{fig_cartoon}(d) returns the electric field to $E_1$. The polarization of the single spin along [001] can now be measured optically, e.g. by Faraday or Kerr rotation\cite{Leuenberger2005b,Mikkelsen2007}.

The two perpendicular spin precession vectors shown in Fig.~\ref{fig_cartoon}(bc) are central to achieving full Bloch sphere control of the spin. The condition that two spin precession vectors are perpendicular, ${\bf \Omega}(E_2) \cdot {\bf \Omega}(E_3)=0$, can be rewritten using Eq.~(\ref{eqn_zeeman}) as
\begin{equation}
\label{eqn_orthogonality}
0= {\rm g}_{\rm [001]}(E_2){\rm g}_{\rm [001]}(E_3) B_{\rm [001]}^2 + {\rm g}_{\rm [110]}(E_2){\rm g}_{\rm [110]}(E_3)B_{\rm [110]}^2
\end{equation}
for $B_{\rm [1\overline{1}0]}=0$.
Equation~(\ref{eqn_orthogonality}) can only be satisfied if one (and only one) of the ${\bf g}$ tensor
components changes sign, which should be ${\rm g}_{\rm [110]}$ to permit the desired configuration in Fig.~\ref{fig_cartoon}(a). For ${\bf B}$ pointing in an arbitrary direction either one or two components of {\bf g} must change sign. A change in sign for a component of {\bf g} has been demonstrated experimentally in quantum wells\cite{Salis2001}, but the observed changes of {\bf g} in quantum dots have been very small\cite{Nakaoka2007}. A change in sign with {\it magnetic} field has been predicted theoretically for large quantum dots\cite{Destefani2005}. Here we rely on a change in electric field to change the sign, which is suggested by the height-dependence of the in-plane components of the {\bf g} tensor\cite{Pryor2006}. This effect was also hinted at by the difference in sign between growth-direction {\bf g} tensor components calculated for spherical and semispherical CdTe dots\cite{Prado2004b}.
Thus we find that to achieve full control over a spin with only a single, growth-direction electric field, as shown in Fig.~\ref{fig_cartoon}(bc), one component of the {\bf g}~tensor must change sign. That component would be preferably along an axis perpendicular to the growth direction. 

We now describe our calculated results of {\bf g} for a broad range of InAs/GaAs quantum dots, and find many dots for which one in-plane component of the ${\bf g}$ tensor changes sign with applied electric field. Our evaluation of the precession frequency of the spin in these dots under the conditions of Fig.~\ref{fig_cartoon}(bc) yields $\sim 150$MHz in a 1 Tesla field. The precession frequency scales with the magnitude of the applied magnetic field, so a reasonable laboratory field can produce a precession frequency in excess of 1~GHz.

We calculate the electron {\bf g}~tensor in the dots using a strain-dependent 8-band ${\bf k}\cdot{\bf p}$ model\cite{Bahder1990} solved in real space\cite{Pryor1998} that includes both the orbital and the spin effects of the magnetic field in a lattice gauge theory formulation\cite{Pryor2006}.  To calculate a {\bf g} tensor component along a principal axis ([001], [110], or [1$\bar{1}$0]), a 1 Tesla field is applied along the corresponding axis and the energy splitting between the two lowest-energy spin-dependent conduction states is determined.  The sign of  each {\bf g} tensor component was found by computing \hbox{$< \psi | {\bf S}\cdot \hat B | \psi >$ }to determine the spin alignment.

\begin{figure}
\includegraphics[width=\columnwidth]{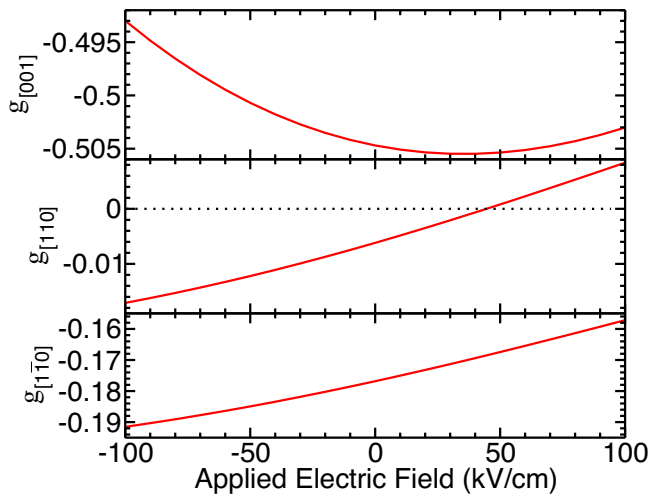}
\caption{{\bf g}~tensor components of a quantum dot as a
  function of an applied electric  field along the [001] direction. The lens-shaped quantum dot has a height of  6.2~nm, base diameter in the [110] direction of 15.6~{\rm nm} and base diameter in the $[1\bar{1}0]$ direction of 9.2~{\rm nm}.  ${\rm g}_{[110]}$ crosses from negative to positive near 50kV/cm.
\label{fig_efield}}
\end{figure}

The electric-field dependence of the {\bf g} tensor components is shown in Fig.~\ref{fig_efield} for a dot with a geometry favorable for spin manipulation for the magnetic field shown in Fig.~\ref{fig_cartoon}. 
The lens-shaped dot has a height of 6.2~nm, with a base elongated along the [110] direction so that the dot diameters in the in-plane directions differ: $d_{[110]}=15.6~{\rm nm}$ and $d_{[1\bar{1}0]} = 9.2~{\rm nm}$. Dot height-to-base aspect ratios and elongations such as this can be achieved by manipulating growth conditions. Even larger height-to-base aspect ratios are possible using pillaring techniques\cite{He2007}.
A key feature of this dot is that ${\rm g}_{[110]}=0$ for an electric field ($\sim 50$~kV/cm) considerably below the breakdown voltage of the host semiconductor ($> 200$~kV/cm). As the dot shape is not symmetrical along the [001] direction the response to the electric field is not symmetric around $ E=0$, and has a sizable $E$-linear dependence. This effect will occur for all asymmetric dots, including lithographic dots, but to a different degree. All electric fields considered in Fig.~\ref{fig_efield} are less than half the typical breakdown field of  GaAs.  The sign of all the {\bf g} tensor components is negative at ${ E}=0$. We can select from Fig.~\ref{fig_efield} the values of the electric field that correspond to the panels in Fig.~\ref{fig_cartoon}. Fig.~\ref{fig_cartoon}(a) and (d) correspond to $E_1=50$kV/cm, for which ${\rm g}_{\rm [110]}=0$ and $\Omega$ is parallel to $ [001]$. Fig.~\ref{fig_cartoon}(b) corresponds to $E_2=-50$kV/cm, for which ${\rm g}_{\rm [110]}=-0.01$. Fig.~\ref{fig_cartoon}(c) corresponds to $E_3=100$kV/cm, for which ${\rm g}_{\rm [110]}=+0.01$. For a magnetic field rotated slightly (3.4$^{\rm o}$) from [110] towards the [001] direction, the spin precession axis will point as shown in Fig.~\ref{fig_cartoon}(bc) for $E_2$ and $E_3$.

\begin{figure}
\includegraphics[width=\columnwidth]{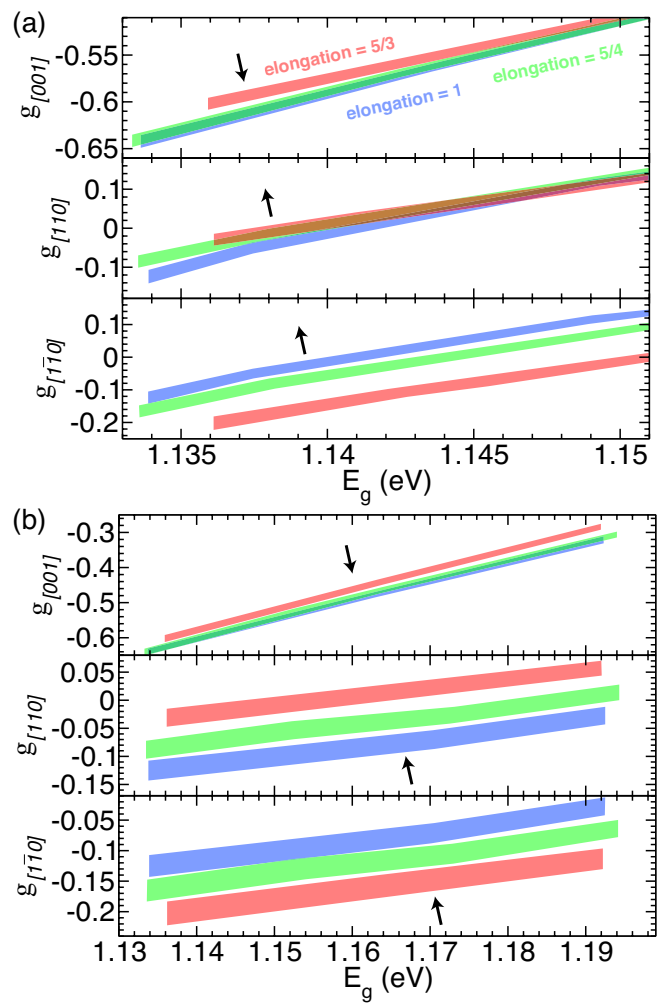}
\caption{
{\bf g}~tensor components for dots of different shapes and sizes, plotted as a function of the resulting energy gap of the dot. Each band corresponds to a fixed footprint shape, characterized by the elongation ($=d_{[110]}/d_{[1\bar{1}0]}$). 
(a) Height is varied  for elongations of 5/3, 5/4, and 1, with a fixed footprint, corresponding to a 13~nm geometric mean of $d_{[110]}$ and $d_{[1\bar{1}0]}$.  (b) Footprint size is varied, with height fixed at 6.2~nm. The colors indicate the elongation of the dot (blue is 1, green is 5/4, red is 5/3). Application of an electric field (from $-100$kV$/$cm to
  $+100$kV$/$cm) modifies the g-factors within the color bands. The change within the band as the electric field is increased is indicated by the arrow in each panel.
\label{combined}}
\end{figure}

Figure~\ref{combined} shows the results of a more extensive study as a function of dot shape and size. For $E=0$ the trends in {\bf g} are similar to those reported earlier for both growth-direction\cite{Nakaoka2004,Pryor2006} and in-plane components\cite{Pryor2006}. In Fig.~\ref{combined}(a) the dot footprint is fixed and the height is changed. The results are plotted as a function of quantum dot energy gap. Fig.~\ref{combined}(b) shows similar results, but the height is fixed and the footprint is changed. The width of the bands corresponds to the change in {\bf g} as the electric field is changed from -100kV/cm to 100kV/cm. As shown in Fig.~\ref{combined}(a), for all three elongations ($=d_{[110]}/d_{[1\bar{1}0]}$) there is a height, corresponding to a quantum dot energy gap of $\sim 1.14$eV, for which ${\rm g}_{\rm [110]}$ will change sign with applied electric field. As shown in Fig.~\ref{combined}(b), however, for fixed height and varying footprint  ${\rm g}_{\rm [110]}$ will change sign with applied electric field for dots with gaps that vary by over 50~meV, permitting some tuning of the optical excitation energy of an optimized dot.

In all cases the smaller dots have larger quantum dot energy gaps, due to the increased confinement of both electrons and holes. Increased electron confinement quenches the orbital angular momentum of the electronic wavefunction\cite{Pryor2006} and drives the {\bf g} tensor closer to the free electron value of 2. Thus dots with larger $E_g$'s will have {\bf g} tensors closer to 2, a trend visible in Fig.~\ref{combined}. A positive electric field ($E_1$ and $E_3$ in Fig.~\ref{fig_cartoon})  pushes the electronic wave function towards the base of the dot, whereas a negative electric field ($E_2$ in Fig.~\ref{fig_cartoon}) pushes it towards the apex. The electric field dependence of the {\bf g} tensor comes from several effects: 
confinement-induced angular momentum quenching\cite{Pryor2006}, changes in the dot energy gap due to the quantum confined Stark effect\cite{Empedocles1997}, and variation in the heavy/light hole spitting from the electron wave function overlapping different parts of the inhomogeneously strained dot.
These effects, along with the anisotropy of the dot confining potential, produce anisotropy in the electric field dependence of the {\bf g} tensor that is even more extreme than the anisotropy in the magnitude of {\bf g}.


We also find that dots with {\bf g} tensor components that change sign are well-suited for resonant oscillation of the {\bf g} tensor ({\bf g} tensor modulation resonance\cite{Kato2003}). Here the approach would be to use a static electric field and add a harmonically oscillating electric field. We linearize the field dependence of the components of the  {\bf g}~tensor,
\begin{equation}
  {\rm g}_i(t) = {\rm g}_{i0} + \frac{\partial  {\rm g}_i}{\partial E} E \cos(\omega t),
\end{equation}
and the resulting precession axis,
\begin{align}
{\bf \Omega}(t) &= {\bf \Omega_0} +
  \cos(\omega t) {\bf \Omega_t}.
\end{align}
 The optimal choice of magnetic field orientation for resonant manipulation corresponds to that field orientation where the component of ${\bf \Omega_t}$ transverse to ${\bf \Omega_0}$ is maximized. The Rabi frequency is
\begin{align}
  \Omega_{Rabi} &= \frac {\mu_B |{\bf \Omega_t}|}{4 \hbar} \left(1 -
  \left( \hat \Omega_t \cdot \hat \Omega_0 \right)^2 \right)^{\frac{1}{2}}
\end{align}
To evaluate the Rabi frequency we neglect the dependence of ${\rm g}_{[001]}$ on electric field over the range shown in Fig.~\ref{fig_efield}, as that dependence is much smaller than the relative changes of the in-plane components. We find that the largest Rabi frequencies can be achieved for any in-plane orientation of the magnetic field if the proper polar angle (measured from [001]) for {\bf B}  is chosen. For the dot characterized in Fig.~\ref{fig_efield}, the optimal polar angle ranges from $\pi/2$ for an in-plane magnetic field parallel to [110] to $\approx \pi/2\pm \pi/6$ for an in-plane magnetic field parallel to [1$\bar{1}$0]. The resulting Rabi frequency is approximately 150~MHz for a 1 Tesla magnetic field, essentially the same as the spin precession frequency obtained for the geometry shown in Fig.~\ref{fig_cartoon}(bc). Thus for a reasonable laboratory magnetic field of 10 Tesla the Rabi frequency would exceed 1 GHz.

Full control of a localized spin using a single growth-direction electric field applied using a nanoscale top gate should provide scalable methods of manipulating single spins in semiconductor quantum dot systems. The ability to set the electric field to make ${\rm g}_{[110]}$ vanish also allows the spin precession vector to be oriented parallel to the growth direction when desired, and permits the convenient storage of optically-excited polarized spins until they are ready for manipulation. 

We acknowledge support from an ONR MURI and an NSF NIRT. 


\begin{thebibliography}{26}
\expandafter\ifx\csname natexlab\endcsname\relax\def\natexlab#1{#1}\fi
\expandafter\ifx\csname bibnamefont\endcsname\relax
  \def\bibnamefont#1{#1}\fi
\expandafter\ifx\csname bibfnamefont\endcsname\relax
  \def\bibfnamefont#1{#1}\fi
\expandafter\ifx\csname citenamefont\endcsname\relax
  \def\citenamefont#1{#1}\fi
\expandafter\ifx\csname url\endcsname\relax
  \def\url#1{\texttt{#1}}\fi
\expandafter\ifx\csname urlprefix\endcsname\relax\def\urlprefix{URL }\fi
\providecommand{\bibinfo}[2]{#2}
\providecommand{\eprint}[2][]{\url{#2}}

\bibitem[{\citenamefont{Awschalom et~al.}(2002)\citenamefont{Awschalom,
  Samarth, and Loss}}]{Awschalom2002}
\bibinfo{editor}{\bibfnamefont{D.~D.} \bibnamefont{Awschalom}},
  \bibinfo{editor}{\bibfnamefont{N.}~\bibnamefont{Samarth}}, \bibnamefont{and}
  \bibinfo{editor}{\bibfnamefont{D.}~\bibnamefont{Loss}}, eds.,
  \emph{\bibinfo{title}{Semiconductor Spintronics and Quantum Computation}}
  (\bibinfo{publisher}{Springer Verlag}, \bibinfo{address}{Heidelberg},
  \bibinfo{year}{2002}).

\bibitem[{\citenamefont{Petta et~al.}(2005)\citenamefont{Petta, Johnson,
  Taylor, Laird, Yacoby, Lukin, Marcus, Hanson, and Gossard}}]{Petta2005}
\bibinfo{author}{\bibfnamefont{J.~R.} \bibnamefont{Petta}},
  \bibinfo{author}{\bibfnamefont{A.~C.} \bibnamefont{Johnson}},
  \bibinfo{author}{\bibfnamefont{J.~M.} \bibnamefont{Taylor}},
  \bibinfo{author}{\bibfnamefont{E.~A.} \bibnamefont{Laird}},
  \bibinfo{author}{\bibfnamefont{A.}~\bibnamefont{Yacoby}},
  \bibinfo{author}{\bibfnamefont{M.~D.} \bibnamefont{Lukin}},
  \bibinfo{author}{\bibfnamefont{C.~M.} \bibnamefont{Marcus}},
  \bibinfo{author}{\bibfnamefont{M.~P.} \bibnamefont{Hanson}},
  \bibnamefont{and} \bibinfo{author}{\bibfnamefont{A.~C.}
  \bibnamefont{Gossard}}, \bibinfo{journal}{Science}
  \textbf{\bibinfo{volume}{309}}, \bibinfo{pages}{2180} (\bibinfo{year}{2005}).

\bibitem[{\citenamefont{Tokura et~al.}(2006)\citenamefont{Tokura, van~der Wiel,
  Obata, and Tarucha}}]{Tokura2006}
\bibinfo{author}{\bibfnamefont{Y.}~\bibnamefont{Tokura}},
  \bibinfo{author}{\bibfnamefont{W.~G.} \bibnamefont{van~der Wiel}},
  \bibinfo{author}{\bibfnamefont{T.}~\bibnamefont{Obata}}, \bibnamefont{and}
  \bibinfo{author}{\bibfnamefont{S.}~\bibnamefont{Tarucha}},
  \bibinfo{journal}{\prl} \textbf{\bibinfo{volume}{96}},
  \bibinfo{pages}{047202} (\bibinfo{year}{2006}).

\bibitem[{\citenamefont{Kalevich and Korenev}(1993)}]{Kalevich1993}
\bibinfo{author}{\bibfnamefont{V.~K.} \bibnamefont{Kalevich}} \bibnamefont{and}
  \bibinfo{author}{\bibfnamefont{V.~L.} \bibnamefont{Korenev}},
  \bibinfo{journal}{JETP Lett.} \textbf{\bibinfo{volume}{57}},
  \bibinfo{pages}{571} (\bibinfo{year}{1993}).

\bibitem[{\citenamefont{Oestreich et~al.}(1996)\citenamefont{Oestreich,
  Hallstein, and R\"uhle}}]{Oestreich1996b}
\bibinfo{author}{\bibfnamefont{M.}~\bibnamefont{Oestreich}},
  \bibinfo{author}{\bibfnamefont{S.}~\bibnamefont{Hallstein}},
  \bibnamefont{and} \bibinfo{author}{\bibfnamefont{W.~W.}
  \bibnamefont{R\"uhle}}, \bibinfo{journal}{IEEE J. Quant. Electron.}
  \textbf{\bibinfo{volume}{2}}, \bibinfo{pages}{747} (\bibinfo{year}{1996}).

\bibitem[{\citenamefont{Ivchenko et~al.}(1997)\citenamefont{Ivchenko, Kiselev,
  and Willander}}]{Ivchenko1997b}
\bibinfo{author}{\bibfnamefont{E.~L.} \bibnamefont{Ivchenko}},
  \bibinfo{author}{\bibfnamefont{A.~A.} \bibnamefont{Kiselev}},
  \bibnamefont{and}
  \bibinfo{author}{\bibfnamefont{M.}~\bibnamefont{Willander}},
  \bibinfo{journal}{Solid State Comm.} \textbf{\bibinfo{volume}{102}},
  \bibinfo{pages}{375} (\bibinfo{year}{1997}).

\bibitem[{\citenamefont{Jiang and Yablonovitch}(2001)}]{Jiang2001}
\bibinfo{author}{\bibfnamefont{H.~W.} \bibnamefont{Jiang}} \bibnamefont{and}
  \bibinfo{author}{\bibfnamefont{E.}~\bibnamefont{Yablonovitch}},
  \bibinfo{journal}{Phys. Rev. B} \textbf{\bibinfo{volume}{64}},
  \bibinfo{pages}{041307} (\bibinfo{year}{2001}).

\bibitem[{\citenamefont{Salis et~al.}(2001)\citenamefont{Salis, Kato, Ensslin,
  Driscoll, Gossard, and Awschalom}}]{Salis2001}
\bibinfo{author}{\bibfnamefont{G.}~\bibnamefont{Salis}},
  \bibinfo{author}{\bibfnamefont{Y.}~\bibnamefont{Kato}},
  \bibinfo{author}{\bibfnamefont{K.}~\bibnamefont{Ensslin}},
  \bibinfo{author}{\bibfnamefont{D.~C.} \bibnamefont{Driscoll}},
  \bibinfo{author}{\bibfnamefont{A.~C.} \bibnamefont{Gossard}},
  \bibnamefont{and} \bibinfo{author}{\bibfnamefont{D.~D.}
  \bibnamefont{Awschalom}}, \bibinfo{journal}{Nature}
  \textbf{\bibinfo{volume}{414}}, \bibinfo{pages}{619} (\bibinfo{year}{2001}).

\bibitem[{\citenamefont{Kato et~al.}(2003)\citenamefont{Kato, Myers, Gossard,
  Levy, and Awschalom}}]{Kato2003}
\bibinfo{author}{\bibfnamefont{Y.}~\bibnamefont{Kato}},
  \bibinfo{author}{\bibfnamefont{R.~C.} \bibnamefont{Myers}},
  \bibinfo{author}{\bibfnamefont{A.~C.} \bibnamefont{Gossard}},
  \bibinfo{author}{\bibfnamefont{J.}~\bibnamefont{Levy}}, \bibnamefont{and}
  \bibinfo{author}{\bibfnamefont{D.~D.} \bibnamefont{Awschalom}},
  \bibinfo{journal}{Science} \textbf{\bibinfo{volume}{299}},
  \bibinfo{pages}{1201} (\bibinfo{year}{2003}).

\bibitem[{\citenamefont{Doty et~al.}(2006)\citenamefont{Doty, Scheibner,
  Ponomarev, Stinaff, Bracker, Korenev, Reinecke, and Gammon}}]{Doty2006}
\bibinfo{author}{\bibfnamefont{M.~F.} \bibnamefont{Doty}},
  \bibinfo{author}{\bibfnamefont{M.}~\bibnamefont{Scheibner}},
  \bibinfo{author}{\bibfnamefont{I.~V.} \bibnamefont{Ponomarev}},
  \bibinfo{author}{\bibfnamefont{E.~A.} \bibnamefont{Stinaff}},
  \bibinfo{author}{\bibfnamefont{A.~S.} \bibnamefont{Bracker}},
  \bibinfo{author}{\bibfnamefont{V.~L.} \bibnamefont{Korenev}},
  \bibinfo{author}{\bibfnamefont{T.~L.} \bibnamefont{Reinecke}},
  \bibnamefont{and} \bibinfo{author}{\bibfnamefont{D.}~\bibnamefont{Gammon}},
  \bibinfo{journal}{\prl} \textbf{\bibinfo{volume}{97}}, \bibinfo{eid}{197202}
  (\bibinfo{year}{2006}).

\bibitem[{\citenamefont{Nakaoka et~al.}(2007)\citenamefont{Nakaoka, Tarucha,
  and Arakawa}}]{Nakaoka2007}
\bibinfo{author}{\bibfnamefont{T.}~\bibnamefont{Nakaoka}},
  \bibinfo{author}{\bibfnamefont{S.}~\bibnamefont{Tarucha}}, \bibnamefont{and}
  \bibinfo{author}{\bibfnamefont{Y.}~\bibnamefont{Arakawa}},
  \bibinfo{journal}{Phys. Rev. B} \textbf{\bibinfo{volume}{76}},
  \bibinfo{eid}{041301} (\bibinfo{year}{2007}).

\bibitem[{\citenamefont{Roth et~al.}(1959)\citenamefont{Roth, Lax, and
  Zwerdling}}]{Roth1959}
\bibinfo{author}{\bibfnamefont{L.~M.} \bibnamefont{Roth}},
  \bibinfo{author}{\bibfnamefont{B.}~\bibnamefont{Lax}}, \bibnamefont{and}
  \bibinfo{author}{\bibfnamefont{S.}~\bibnamefont{Zwerdling}},
  \bibinfo{journal}{Phys. Rev.} \textbf{\bibinfo{volume}{114}},
  \bibinfo{pages}{90} (\bibinfo{year}{1959}).

\bibitem[{\citenamefont{Ivchenko and Pikus}(1997)}]{Ivchenko1997}
\bibinfo{author}{\bibfnamefont{E.~L.} \bibnamefont{Ivchenko}} \bibnamefont{and}
  \bibinfo{author}{\bibfnamefont{G.~E.} \bibnamefont{Pikus}},
  \emph{\bibinfo{title}{Superlattices and Other Heterostructures}}
  (\bibinfo{publisher}{Springer}, \bibinfo{address}{New York},
  \bibinfo{year}{1997}).

\bibitem[{\citenamefont{Pryor and Flatt\'e}(2006)}]{Pryor2006}
\bibinfo{author}{\bibfnamefont{C.~E.} \bibnamefont{Pryor}} \bibnamefont{and}
  \bibinfo{author}{\bibfnamefont{M.~E.} \bibnamefont{Flatt\'e}},
  \bibinfo{journal}{\apl} \textbf{\bibinfo{volume}{88}},
  \bibinfo{pages}{233108} (\bibinfo{year}{2006}).

\bibitem[{\citenamefont{Meier and Zachachrenya}(1984)}]{Meier1984}
\bibinfo{author}{\bibfnamefont{F.}~\bibnamefont{Meier}} \bibnamefont{and}
  \bibinfo{author}{\bibfnamefont{B.~P.} \bibnamefont{Zachachrenya}},
  \emph{\bibinfo{title}{Optical Orientation: Modern Problems in Condensed
  Matter Science}}, vol.~\bibinfo{volume}{8}
  (\bibinfo{publisher}{North-Holland}, \bibinfo{address}{Amsterdam},
  \bibinfo{year}{1984}).

\bibitem[{\citenamefont{Kroutvar et~al.}(2004)\citenamefont{Kroutvar, Ducommun,
  Heiss, Bichler, Schuh, Abstreiter, and Finley}}]{Kroutvar2004}
\bibinfo{author}{\bibfnamefont{M.}~\bibnamefont{Kroutvar}},
  \bibinfo{author}{\bibfnamefont{Y.}~\bibnamefont{Ducommun}},
  \bibinfo{author}{\bibfnamefont{D.}~\bibnamefont{Heiss}},
  \bibinfo{author}{\bibfnamefont{M.}~\bibnamefont{Bichler}},
  \bibinfo{author}{\bibfnamefont{D.}~\bibnamefont{Schuh}},
  \bibinfo{author}{\bibfnamefont{G.}~\bibnamefont{Abstreiter}},
  \bibnamefont{and} \bibinfo{author}{\bibfnamefont{J.~J.}
  \bibnamefont{Finley}}, \bibinfo{journal}{Nature}
  \textbf{\bibinfo{volume}{432}}, \bibinfo{pages}{81} (\bibinfo{year}{2004}).

\bibitem[{\citenamefont{Pryor and Flatt\'e}(2003)}]{Pryor2003}
\bibinfo{author}{\bibfnamefont{C.~E.} \bibnamefont{Pryor}} \bibnamefont{and}
  \bibinfo{author}{\bibfnamefont{M.~E.} \bibnamefont{Flatt\'e}},
  \bibinfo{journal}{\prl} \textbf{\bibinfo{volume}{91}}, \bibinfo{eid}{257901}
  (\bibinfo{year}{2003}).

\bibitem[{\citenamefont{Leuenberger et~al.}(2005)\citenamefont{Leuenberger,
  Flatt\'e, and Awschalom}}]{Leuenberger2005b}
\bibinfo{author}{\bibfnamefont{M.~N.} \bibnamefont{Leuenberger}},
  \bibinfo{author}{\bibfnamefont{M.~E.} \bibnamefont{Flatt\'e}},
  \bibnamefont{and} \bibinfo{author}{\bibfnamefont{D.~D.}
  \bibnamefont{Awschalom}}, \bibinfo{journal}{Phys. Rev. Lett.}
  \textbf{\bibinfo{volume}{94}}, \bibinfo{pages}{107401}
  (\bibinfo{year}{2005}).

\bibitem[{\citenamefont{Mikkelsen et~al.}(2007)\citenamefont{Mikkelsen,
  Berezovsky, Stoltz, Coldren, and Awschalom}}]{Mikkelsen2007}
\bibinfo{author}{\bibfnamefont{M.~H.} \bibnamefont{Mikkelsen}},
  \bibinfo{author}{\bibfnamefont{J.}~\bibnamefont{Berezovsky}},
  \bibinfo{author}{\bibfnamefont{N.~G.} \bibnamefont{Stoltz}},
  \bibinfo{author}{\bibfnamefont{L.~A.} \bibnamefont{Coldren}},
  \bibnamefont{and} \bibinfo{author}{\bibfnamefont{D.~D.}
  \bibnamefont{Awschalom}}, \bibinfo{journal}{Nature Physics}
  \textbf{\bibinfo{volume}{3}}, \bibinfo{pages}{770} (\bibinfo{year}{2007}).

\bibitem[{\citenamefont{Destefani and Ulloa}(2005)}]{Destefani2005}
\bibinfo{author}{\bibfnamefont{C.~F.} \bibnamefont{Destefani}}
  \bibnamefont{and} \bibinfo{author}{\bibfnamefont{S.~E.} \bibnamefont{Ulloa}},
  \bibinfo{journal}{Phys. Rev. B} \textbf{\bibinfo{volume}{71}},
  \bibinfo{eid}{161303} (\bibinfo{year}{2005}).

\bibitem[{\citenamefont{Prado et~al.}(2004)\citenamefont{Prado, Trallero-Giner,
  Alcalde, L\'opez-Richard, and Marques}}]{Prado2004b}
\bibinfo{author}{\bibfnamefont{S.~J.} \bibnamefont{Prado}},
  \bibinfo{author}{\bibfnamefont{C.}~\bibnamefont{Trallero-Giner}},
  \bibinfo{author}{\bibfnamefont{A.~M.} \bibnamefont{Alcalde}},
  \bibinfo{author}{\bibfnamefont{V.}~\bibnamefont{L\'opez-Richard}},
  \bibnamefont{and} \bibinfo{author}{\bibfnamefont{G.~E.}
  \bibnamefont{Marques}}, \bibinfo{journal}{Phys. Rev. B}
  \textbf{\bibinfo{volume}{69}}, \bibinfo{pages}{201310R}
  (\bibinfo{year}{2004}).

\bibitem[{\citenamefont{Bahder}(1990)}]{Bahder1990}
\bibinfo{author}{\bibfnamefont{T.~B.} \bibnamefont{Bahder}},
  \bibinfo{journal}{Phys. Rev. B} \textbf{\bibinfo{volume}{41}},
  \bibinfo{pages}{11992} (\bibinfo{year}{1990}).

\bibitem[{\citenamefont{Pryor}(1998)}]{Pryor1998}
\bibinfo{author}{\bibfnamefont{C.}~\bibnamefont{Pryor}},
  \bibinfo{journal}{Phys. Rev. B} \textbf{\bibinfo{volume}{57}},
  \bibinfo{pages}{7190} (\bibinfo{year}{1998}).

\bibitem[{\citenamefont{He et~al.}(2007)\citenamefont{He, Krenner, Pryor,
  Zhang, Wu, Allen, Morris, Sherwin, and Petroff}}]{He2007}
\bibinfo{author}{\bibfnamefont{J.}~\bibnamefont{He}},
  \bibinfo{author}{\bibfnamefont{H.}~\bibnamefont{Krenner}},
  \bibinfo{author}{\bibfnamefont{C.}~\bibnamefont{Pryor}},
  \bibinfo{author}{\bibfnamefont{J.~P.} \bibnamefont{Zhang}},
  \bibinfo{author}{\bibfnamefont{Y.}~\bibnamefont{Wu}},
  \bibinfo{author}{\bibfnamefont{D.}~\bibnamefont{Allen}},
  \bibinfo{author}{\bibfnamefont{C.}~\bibnamefont{Morris}},
  \bibinfo{author}{\bibfnamefont{M.}~\bibnamefont{Sherwin}}, \bibnamefont{and}
  \bibinfo{author}{\bibfnamefont{P.~M.} \bibnamefont{Petroff}},
  \bibinfo{journal}{Nano Letters} \textbf{\bibinfo{volume}{7}},
  \bibinfo{pages}{802} (\bibinfo{year}{2007}).

\bibitem[{\citenamefont{Nakaoka et~al.}(2004)\citenamefont{Nakaoka, Saito,
  Tatebayashi, and Arakawa}}]{Nakaoka2004}
\bibinfo{author}{\bibfnamefont{T.}~\bibnamefont{Nakaoka}},
  \bibinfo{author}{\bibfnamefont{T.}~\bibnamefont{Saito}},
  \bibinfo{author}{\bibfnamefont{J.}~\bibnamefont{Tatebayashi}},
  \bibnamefont{and} \bibinfo{author}{\bibfnamefont{Y.}~\bibnamefont{Arakawa}},
  \bibinfo{journal}{Phys. Rev. B} \textbf{\bibinfo{volume}{70}},
  \bibinfo{pages}{235337} (\bibinfo{year}{2004}).

\bibitem[{\citenamefont{Empedocles and Bawendi}(1997)}]{Empedocles1997}
\bibinfo{author}{\bibfnamefont{S.~A.} \bibnamefont{Empedocles}}
  \bibnamefont{and} \bibinfo{author}{\bibfnamefont{M.~G.}
  \bibnamefont{Bawendi}}, \bibinfo{journal}{Science}
  \textbf{\bibinfo{volume}{278}}, \bibinfo{pages}{2114} (\bibinfo{year}{1997}).

\end{thebibliography}

\end{document}